\documentstyle[aps,prc,psfig]{revtex}
 
\begin{document}

\title{
Isoscalar $NN$ spin-orbit potential from a Skyrme model
with scalar mesons\thanks{Supported by the EPSRC, UK.}}

\author{ Abdellatif  Abada\thanks{
Address after Feb. $1^{\rm st}$, 1997: BP Finance, 
BP International Ltd, Britanic House, 1 Finsbury Circus, London EC2M 7BA.}} 

\address{
 Theoretical Physics Group, Department of Physics and Astronomy,\\
  University of Manchester, Manchester M13 9PL, England }

\date{\today}

\maketitle

\begin{abstract}
As a first step toward circumventing the difficulty to obtain an
attractive isospin-independent $NN$ spin-orbit force from Skyrme-type
models involving only pions, we investigate an improved Skyrme Lagrangian
that incorporates the scalar-isoscalar meson $\epsilon$ which can be
viewed as the cause behind the enhancement of the $\pi \pi\
S$-wave. We find that at large distances, the main contribution to
the spin-orbit potential comes from the scalar Lagrangian and it
is found to be attractive. We briefly discuss how to pursue
this work to finally obtain a medium-range attractive interaction. 

\vskip 0.5cm
\noindent PACS numbers(s): 11.10 Lm, 12.39 Dc, 13.75 Cs. 

\noindent Report-no: MC/TH 96/17

\noindent hep-ph/9606270
\end{abstract}

\section{INTRODUCTION}

Ten years before the advent of QCD, Skyrme proposed a model~\cite{Sk}
describing hadronic physics which involves only pion fields and where
baryons emerge as topological solitons. This model is recognized as the
simplest chiral realization of QCD at low energies and large $N_c$~\cite{Wi}.
The corresponding Lagrangian contains, in addition to the well known
nonlinear $\sigma$ model, an antisymmetric term of fourth order in powers of
the derivatives of the pion field (the so-called Skyrme term).
The latter has been added by Skyrme in order to avoid soliton collapse.
Despite its relative success in describing some properties of baryons
({\it cf.} Ref.~\cite{ZB} for a review), the Skyrme model presents several 
shortcomings such as its prediction of a very small value for the axial
coupling constant~\cite{ANW} and an almost zero nucleon mass when 
the Casimir effects are taken into account~\cite{Mou}. But probably
the most important crisis of the Skyrme model is that it does not
allow for the formation of nuclei. Indeed, it predicts a repulsive $NN$
central potential~\cite{JJP}. Because of these drawbacks, several
extensions of the standard Skyrme model have been proposed in order to
improve its prediction power. The most famous one consists in adding
to the Skyrme Lagrangian a term of order six in powers of the derivatives
of the pion field, proportional to the square of the baryon current 
as first proposed in~\cite{Jac}. The extended Skyrme model thus obtained 
contains a second, fourth and sixth-order term, but still is an effective 
theory of pions. It has been shown recently~\cite{Mou,AM} that
this extended model is much more realistic when describing low-energy
hadronic physics than the standard Skyrme model. 

In contrast with the velocity-independent part of the $NN$ potential, 
the kinetic part and especially the $NN$ spin-orbit force has not been
studied extensively within the framework of Skyrme-type models.
However, all the authors who worked on this subject, e.g.~\cite{autho,WA},
arrived at the disappointing result that the standard Skyrme model predicts
a repulsive isospin-independent spin-orbit force while, according to
the phenomenology~\cite{PB}, it should be attractive. A glimpse of hope
has emerged from Ref.~\cite{RS} where it was claimed that when considering
the sixth-order term, an attractive isoscalar spin-orbit interaction could
be obtained. However, we have shown recently~\cite{Ab2} that this is not the
case: the sixth-order term contributes with a positive sign as is the case
for the Skyrme term. Indeed, the authors of Ref.~\cite{RS} considered in
their calculations only one piece of the force due to the sixth-order term
and omitted the piece that stems from the baryon exchange current,
which turns out to be not only repulsive, but the dominant one 
too~\cite{Ab2}. 
Thus, despite its success in improving the predictions of 1-baryon data,
even an extended Skyrme model including higher-order terms
in powers of the derivatives of the pion field fails to predict an
attractive $NN$ isoscalar spin-orbit interaction as it fails to 
predict the correct $NN$ central potential~\cite{Ka}. Therefore,
one has to consider more realistic low-energy models to
describe the $NN$ interaction.

The Skyrme term as well as the sixth-order term can both be regarded as an
{\it infinite} mass limit, the so-called local approximation, of a model
with $\rho$-meson~\cite{Ike} and $\omega$-meson~\cite{Jac} exchanges
respectively. These findings have provided the ground for improving
Skyrme-type models by constructing effective Lagrangians which include,
in addition to the pions, the other low-lying mesons. The authors of 
Ref.~\cite{KV} have investigated the $NN$ force by considering such an 
effective
model incorporating the mesons $\pi, \rho, \omega, A_1$ and the scalar
meson $\epsilon$. Unlike previous works, they have introduced the vector
meson fields by gauging the linear $\sigma$-model instead of the nonlinear
one. They have found that the different channels of the $NN$
interaction arising from the potential piece of the Lagrangian are very well
described, in particular the central potential for which an attraction has
been found in a quantitative agreement with the phenomenology~\cite{PB}.
These results lead us to think that, similarly to the central potential,
it may be that the right way to circumvent the difficulty of finding an
attractive isoscalar spin-orbit force with Skyrme-type models is to consider
effective theories which incorporate low-mass mesons with {\it finite} 
masses.
In order to check these claims, one has to investigate the kinetic energy of
a two-nucleon system within an effective Lagrangian such as the one 
in~\cite{KV}. This is of course not an easy task and goes beyond the scope of
this short paper. As a first step, we rather focus here only on the scalar
meson contribution to the isoscalar spin-orbit force. Indeed, it has been
pointed out in Ref.~\cite{KV} that the presence of scalar degrees of freedom
plays the crucial role of recovering the missed central attraction, so one
may think that this will also be the case for the isoscalar spin-orbit force.

This idea to account for a scalar field in the spin-orbit channel has been 
investigated in Ref.~\cite{KE}. There, a dilaton field is coupled
to Skyrmions in order to mimic the scale breaking of QCD. 
We will not discuss here if 
a dilaton field is suitable or not to provide
a good description of low-energy hadron physics; we refer the interested 
reader to Ref.~\cite{Mk} for a detailed discussion on this subject.
The fact is that a combination of the sixth-order term and dilaton coupling
yields an attractive isoscalar spin-orbit force, as it has been claimed 
in~\cite{KE}. 
Note though that this result remains questionable since
these authors, like those of Ref.~\cite{RS}, ignored
the baryon exchange current contribution in their calculations~\cite{Ab2}.
In any case, here we will follow a more transparent way in order to
incorporate scalar degrees of freedom~\cite{KV}. It consists in considering
the linear $\sigma$ model in which the presence of the scalar field is
intrinsic. In Sec.II, the model with the scalar-isoscalar $\epsilon$-meson
is presented. Sec.III contains the derivation of the isoscalar spin-orbit
force. Our results and discussions are presented in Sec.IV.

\section{THE MODEL WITH SCALAR MESONS}

The infinite-mass limit of the effective Lagrangian with 
 $\pi, \rho, \omega$ and $\epsilon$ mesons~\cite{KV} reads:
\begin{equation} \begin{array} {ll} \label{La}
{\cal L}  &{\displaystyle ={\cal L}_{{\rm NL}\sigma} +{\cal L}_4 +{\cal L}_6  
+{\cal L}_{\chi {\rm SB}} }\\
&{\displaystyle =\frac {f_{\pi}^2 }{4}
\hbox {Tr} ~(\partial_{\mu} U \partial^{\mu} U^{+})
+ \frac {1}{32 e^2}
\hbox {Tr} \left[ (\partial_{\mu} U) U^{+} , (\partial_{\nu} U) U^{+}
\right]^2   -  e_6^2
B_{\mu} B^{\mu}  +\frac {f_{\pi}^2  m_{\pi}^2}{2}\hbox {Tr}(U -1)} 
\end{array} \end{equation}
where  $B^{\mu} =  \epsilon ^{\mu \nu \alpha \beta} \hbox {Tr} \left(~
(\partial_{\nu} U) U^+ (\partial_{\alpha} U) U^+ 
(\partial_{\beta} U) U^+~\right) /24 \pi^2$
is the baryon current~\cite{Sk}, and 
$U$ an SU(2) matrix which characterizes the pion field.
The first term in Eq.~(\ref{La}) corresponds to the 
nonlinear $\sigma$ model, $f_{\pi}$ being the pion decay constant. 
The second term, parameterized by the coupling constant $e$,
is the so-called Skyrme term. 
The third term is of order six in the derivatives of the pion field
and corresponds to $\omega$-meson exchange in the case of
an infinite $\omega$-meson mass~\cite{Jac}. The coupling constant
$e_6^2$ is a parameter related to the $\omega \to \pi \gamma$ width.
The last term in Eq.~(\ref{La})
which is proportional to the square of the pion mass $m_{\pi}$  (139 MeV)
implements a small explicit breaking of chiral symmetry. As compared to
the standard Skyrme model, the extended model~(\ref{La}), when used with
realistic parameters, provides a more accurate description of the 1-baryon
properties (e.g., the nucleon quantum mass, the $\Delta$-$N$ mass splitting,
the breathing mode energy the axial-vector coupling constant 
$g_A$,...)~\cite{Mou,AM}. 
However, as mentioned in the introduction, it does not
describe properly neither the $NN$ isospin-independent spin-orbit 
force~\cite{Ab2} nor the central potential~\cite{Ka}.

The scalar-isoscalar meson $\epsilon$ is viewed as the responsible for
the $S$-wave attraction in the $\pi \pi$ interaction
and, therefore, should be considered as an essential ingredient in 
the low-energy hadronic phenomenology. It is incorporated
in the Lagrangian~(\ref{La}) by replacing the nonlinear $\sigma$ model 
${\cal L}_{{\rm NL}\sigma}$ with the linear one. Namely,
\begin{equation}\label{Ls}
\displaystyle {\cal L}_{{\rm NL}\sigma} \to {\cal L}_{{\rm L}\sigma} =
\frac{1}{2} \partial_{\mu}\xi \partial^{\mu}\xi + \frac {\xi^2 }{4} ~
\hbox {Tr} ~(\partial_{\mu} U \partial^{\mu} U^{+}) - 
\lambda(\xi^2-f_{\pi}^2)^2
\end{equation}
where $\xi$ is the isoscalar content of the quaternion field
$\sigma+i\mbox{\boldmath $\tau.\pi$}$, $\sigma$ being the scalar chiral 
partner of the pion field. The scalar field $\epsilon$ is defined as 
$\epsilon({\bf r})=f_{\pi}-\xi({\bf r})$. The coupling constant $\lambda$ is
related to the $\epsilon$-meson mass through 
$m_{\epsilon}^2=8 \lambda f_{\pi}^2$. It is straightforward to check that by 
taking $m_{\epsilon}\to \infty$ in Eq.~(\ref{Ls}), one recovers the 
nonlinear 
$\sigma$ model ${\cal L}_{{\rm NL}\sigma}$. Indeed, in that limit,
$\lambda$ becomes infinite and so the $\xi$-field has to be frozen to
its asymptotic value $f_{\pi}$ in order to keep the potential energy
in Eq.~(\ref{Ls}) finite.

Therefore, for a realistic finite $\epsilon$-meson mass, the Lagrangian
density~(\ref{La}) is then replaced by the following one:
\begin{equation}  \label{Lanew}
{\cal L}'={\cal L}_{{\rm L}\sigma} +{\cal L}_4 +{\cal L}_6 
+{\cal L}_{\chi {\rm SB}}
\end{equation}
The one-soliton system is commonly solved by assuming the hedgehog ansatz 
for the pion field: 
\begin{equation}\label{Uh}
U({\bf r}) \equiv U_H({\bf r}) = \displaystyle 
\exp\left(~i\mbox{\boldmath $\tau$}.\hat {\bf r} F(r) ~\right)
\end{equation}
where the $\tau_a$'s are the Pauli matrices and the notation 
$\hat {\bf r}$ means ${\bf r}/r$. 
The static Euler-Lagrange equations for the 
chiral field $F$ and the $\xi$-field corresponding to the 
Lagrangian~(\ref{Lanew}) read
\begin{equation} \begin{array} {cc}\label{eqs}
\displaystyle 
\left( \left[~r^2\xi^2 + \frac{2}{e^2}\sin^2(F)+\frac{e_6^2}
        {2\pi^4~r^2}\sin^4(F)~\right]~ F'\right)'=
\left(\xi^2+\frac{1}{e^2r^2}\sin^2(F)~\right)\sin(2F) ~+\\
\displaystyle ~~~~~~~~~~~~~~~~~~~~~~~~~~~~~~~~~~~~~~
 \left(\frac{1}{e^2}+\frac{e_6^2}
{2\pi^4~r^2}\sin^2(F)~\right) \sin(2F)~F^{'2}~+
f_{\pi}^2m_{\pi}^2 r^2 \sin(F) ~,\\ 
\displaystyle
(r\xi)''=\left(F^{'2}+\frac{2}{r^2} \sin^2(F) +
4 \lambda(\xi^2-f_{\pi}^2)~\right)r\xi
\end{array}
\end{equation}
where primes indicate radial-coordinate differentiation. 
In order to ensure a winding number one the chiral angle $F$ 
obeys the usual boundary conditions while the $\xi$-field fulfills the 
conditions $\xi'(0)=0$ and $\xi(\infty)=f_{\pi}$. 
Let us observe that the above Euler-Lagrange equations are solved with 
a set of parameters whose values have been fixed by fitting to 
the mesonic sector~\cite{Mou,AM,KV}. Namely, $f_{\pi}=93$~MeV, $e=7.2$, 
$e_6=1.66$~fm, $m_{\epsilon}=650$~MeV. These parameters yield, e.g., 
a value of the nucleon mass
of $\sim 1$~GeV after subtracting the Casimir energy~\cite{Mou}, and a
$\Delta$-$N$ mass splitting of~$\sim$~267 MeV. 
We have plotted in Fig.1 the chiral function $F$ and the 
$\xi$-field, solutions to the equations of motion~(\ref{eqs}) 
obtained with the set of parameters given above. 
We will use below these two functions as an input
in the numerical computation of the spin-orbit force. 
For reference, we have also shown in Fig.1 the chiral solution $F$ in the 
case of the model~(\ref{La}), i.e. $\xi=f_{\pi}$,~\cite{AM}.

\section{THE SPIN-ORBIT FORCE}

So far, all the calculations on the 
$NN$ spin-orbit force within Skyrme-type models have been carried
out in the framework of the product ansatz as suggested by Skyrme~\cite{Sk2}
 ({\it cf.} Ref.~\cite{autho,WA} and references therein). 
It is worthwhile noticing that the latter is only a simple approximation 
to the two-baryon configuration and not a self-consistent solution. 
We will not discuss here the degree of validity of this approximation. 
It is commonly chosen  because of its relative 
simplicity as compared to other two-baryon field configurations which can be 
found in the literature~\cite{NW}. 
Furthermore, it becomes exact for large $NN$ 
separation\footnote{The region of validity of the product ansatz corresponds 
to a relative distance $r$ much larger than 1 fm.}. Thus,
following common practice, we use it here to describe a system of two
interacting solitons. Furthermore, in order to obtain the appropriate
spin and isospin structure, we also introduce rotational 
dynamics~\cite{ANW}. 
Hence, the field configuration of the two-nucleon system separated by 
a vector ${\bf r}$ reads:
\begin{equation} \begin{array}{cc}\label{pro}
U(A_1,A_2,{\bf x}, {\bf r}) = U_1 U_2 =
 A_1 U_H( {\bf r}_1)~A_1^+A_2~U_H( {\bf r}_2) A_2^+ ~~,\\
{\bf r}_1 = {\bf x}-{\bf r}/2 ~~,~~~ 
{\bf r}_2 = {\bf x}+{\bf r}/2~~,
\end{array} \end{equation}
where $A_1$ and $A_2$ are SU(2) matrices,
and $U_H$ the hedgehog single soliton~(\ref{Uh}).
For the scalar field, we use the configuration suggested in Ref.~\cite{KV}
\begin{equation}\label{xi12}
\displaystyle \xi_{\rm P}({\bf x}, {\bf r})=\frac{1}{f_{\pi}}
\xi({\bf r}_1)\xi({\bf r}_2)
\end{equation}
which obviously is the form the most compatible with the product-ansatz
approximation~(\ref{pro}).
To carry out a simultaneous quantization of the relative motion of the two 
nucleons and the rotational motion, we need to treat ${\bf r}, A_1$ and
$A_2$ as collective coordinates. Therefore, we make all these parameters
(${\bf r}, A_1, A_2$) time-dependent.

The spin-orbit potential will emerge from a coupling between the relative
motion and the spins of the two nucleons so that we have 
to calculate the kinetic energy corresponding to~(\ref{Ls}). 
As reported in Ref.~\cite{WA,Ab2}, generally, one has to treat with care
the conversion from velocities to canonical momenta
before identifying and extracting the spin-orbit potential. Indeed,
one has to start from a Lagrangian formalism, take its  ``classical''
kinetic energy and extract from it the mass matrix and then
invert it properly in order to move to a Hamiltonian formalism.
However, for a large relative distance $r$, the region of validity of the 
product ansatz, this procedure~\cite{WA} is equivalent to that of 
Refs.~\cite{autho,RS,Ab2} which we will use here. 
In this latter one starts, in the case~(\ref{Ls}), directly from 
\begin{equation} \label{Ks} 
\displaystyle K_{{\rm L}\sigma}(A_1,A_2,{\bf r})= (-1)
\int \hbox {d}^3 x ~
\left(\frac{1}{4}\xi_{\rm P}^2 \hbox{Tr} (\partial_0U\partial_0U^+) + 
\frac{1}{2} (\partial_0\xi_{\rm P})^2 \right)~
\end{equation}
where the minus sign in front of the integral is put in explicitly so that
account is taken of the change in sign of the 
off-diagonal terms of the $2\times 2$ mass matrix under inversion~\cite{WA}.
Afterwards, one makes the usual identifications~\cite{ANW,Nym}:
\begin{equation} \label{subst}
\displaystyle
\dot {\bf r}_n \to \frac{{\bf p}^{(n)}}{M} ~~,~~ 
 \mbox{\boldmath $\omega$}_n = -\frac{i}{2}
\hbox{Tr} ( \mbox{\boldmath $\tau$}
 A_n^+\dot A_n)  \to \frac{{\bf s}^{(n)}}{2\Lambda}~~,~~ n=1,2 ~~,
\end{equation}
where ${\bf p}^{(n)}$ and  ${\bf s}^{(n)}$ are respectively the radial
momentum and the spin of the $n$-th nucleon while $M$ and $\Lambda$ are 
respectively the mass and the moment of inertia of the single soliton.
Inserting the product ans\"atze~(\ref{pro},\ref{xi12}) in the 
expression~(\ref{Ks}) gives
\begin{equation}\label{Ks2}\displaystyle 
 K_{{\rm L}\sigma}= -\frac{1}{2f_{\pi}^2}\int \hbox {d}^3 x ~
\xi^2({\bf r}_1)\xi^2({\bf r}_2) ~(R_{0a}(U_1)+L_{0a}(U_2)~)^2 ~
+\cdots
\end{equation}
where we have omitted to write the second term in Eq.~(\ref{Ks}) since
it does not contain angular velocities and thus will not contribute to
the spin-orbit force. In Eq.~(\ref{Ks2}), $R_{0a}$ and $L_{0a}$ are the
time components of the right and left currents respectively:
\begin{equation} \begin{array} {ll}\label{currents}
\displaystyle
R_{0a} (U_1) = -\frac{i}{2} \hbox {Tr} (\tau_a U_1^+ ~\partial_{0} U_1 ) 
=D_{ab} (A_1) ~\left(T_{cb}^{(1)} ~\frac{s^{(1)}_c}{2\Lambda} + 
\frac{p^{(1)}_i}{M} R_{ib}^{(1)} ~\right) ~,\\
\displaystyle
L_{0a} (U_2) = -\frac{i}{2} \hbox {Tr} (\tau_a \partial_{0} U_2 ~U_2^+ ) 
=D_{ab} (A_2) ~\left(-T_{bc}^{(2)} ~\frac{s^{(2)}_c}{2\Lambda} + 
\frac{p^{(2)}_i}{M} R_{bi}^{(2)} ~\right) ~.
\end{array}\end{equation}
In the above equations, the sum from 1 to 3 on repeated indices is 
understood.
The tensors $T_{ab}^{(n)}$ and $R_{ab}^{(n)}$ depend only on the position 
${\bf r}_n$ of the $n$-th nucleon, and their explicit 
expressions can be found
in Eq.~(11) of Ref.~\cite{Ab2}. The $D_{ab}$'s are the matrix elements of 
the $3\times3$ rotation matrix in the adjoint representation:  
$D_{ab}(A)=\hbox {Tr}~(\tau_a A \tau_b A^+)/2$. To obtain the right-hand
sides of expressions~(\ref{currents}), we have used the quantization 
scheme~(\ref{subst}).
The next step now is to expand the square in the expression of 
$K_{{\rm L}\sigma}$~(\ref{Ks2}) after inserting expressions~(\ref{currents})
and making use of the $D$-matrix properties and
the definitions of the tensors $T_{ab}^{(n)}$ and $R_{ab}^{(n)}$~\cite{Ab2}.
Thus, by considering the isoscalar part\footnote{
Due to the projection theorem, the terms proportional to 
$D(A_1^+A_2)$ contribute to the  isospin dependent force.} and
keeping only terms proportional to ${\bf L.S}$, where 
${\bf S}={\bf s}^{(1)}+{\bf s}^{(2)}$ is the total spin and 
${\bf L}={\bf r}\times {\bf p}$ the angular momentum
(${\bf p}$ being the relative momentum, i.e.,  
${\bf p}={\bf p}^{(2)}=-{\bf p}^{(1)}$), we obtain:
\begin{equation} \begin{array}{ll}\label{spor}
\displaystyle
K_{{\rm L}\sigma} \to \frac{1}{M\Lambda}V_{{\rm L}\sigma}(r)~{\bf L.S}\\
\displaystyle V_{{\rm L}\sigma}(r)=-\frac{1}{f_{\pi}^2~r}
\int {\rm d}^3x~\xi^2(r_1)\xi^2( r_2) 
\frac{1}{r_1} \sin^2(F(r_1)) ~\hat{\bf r}.\hat{\bf r}_1 ~.
\end{array}\end{equation}
The total isoscalar spin-orbit potential of the model~(\ref{Lanew}) is then:
\begin{equation}\label{SO}
\displaystyle
V_{{\rm SO},I=0}=\frac{1}{M\Lambda}\left(V_{{\rm L}\sigma}(r)
+V_4(r)+V_6(r)\right)\bf {L.S}
\end{equation}
where the linear $\sigma$-model contribution $V_{{\rm L}\sigma}$ is given
in Eq.~(\ref{spor}). The  rather lengthy expressions of the Skyrme-term 
contribution $V_4$ and the sixth-order term contribution $V_6$ can be found 
explicitly in  Refs.~\cite{autho,WA,KE} and \cite{Ab2}, respectively, 
and do not need to be rewritten here.
Note that the  analytical expressions 
of $V_4$ and $V_6$ are not affected by the presence of the scalar field
and still depend explicitly only on the chiral field $F$. This is because
the $\xi$-field does not couple to ${\cal L}_4$ nor ${\cal L}_6$, as 
it can be seen from Eq.~(\ref{Lanew}). 
But obviously their numerical values will 
change due to the dependence of $F$ on the scalar degrees of freedom.

\section{RESULTS AND DISCUSSION}

 Before displaying our numerical results concerning $V_{{\rm L}\sigma}$,
let us first discuss the
analytical properties of the terms $V_4(r)$ and $V_6(r)$ at large
distances [{\it cf}. Eq.~(\ref{SO})]. 
As already shown in Ref.~\cite{Ab2}, these 
potentials decrease as $e^{-2m_{\pi}r}$ a large $r$. 
Indeed, one can see from Eq. (\ref{eqs}) that at large distances
the chiral angle $F(r)$ falls off as 
$e^{-m_{\pi}r}/r$ (note that at large $r$, $\xi \to f_{\pi}$).
Then, following the approximation of 
Ref.~\cite{Nym} which consists of treating at large distances the 
field of one soliton as constant in the presence of the other, 
one obtains from~\cite{autho,WA,KE} that $V_4$ decreases as 
$e^{-2m_{\pi}r}$ over a power of $r$. For the potential $V_6$ we have shown
in Ref.~\cite{Ab2} that it is a sum of two terms. 
The first term (the one which has been considered 
in~\cite{RS,KE}) decreases as $e^{-3m_{\pi}r}$.
This behaviour is expected since the sixth-order term ${\cal L}_6$ 
in Eq.~(\ref{La})
can be regarded as the local approximation of an effective model of 
$\omega$ meson coupled to three pions~\cite{Jac}.
However  the asymptotic expression of the second part of $V_6$, 
which arises from the baryon exchange current~\cite{Ab2},
contains both $e^{-2m_{\pi}r}$ and $e^{-3m_{\pi}r}$ 
terms so that it behaves at large $r$ as $e^{-2m_{\pi}r}$. 
This means that the Lagrangian ${\cal L}_6$  generates, 
in addition to the expected three-pion exchange piece, 
a two-pion exchange one (coming from the exchange current)  
similar to that of the the fourth-order Skyrme term ${\cal L}_4$. 
Thus we see that for large distance both potentials $V_4$ and $V_6$ are 
of Yukawa type with range $2m_{\pi}$. In fact, the main motivation of 
including a scalar-isoscalar meson field by hand in Skyrme-type 
models~\cite{KV,KE} is to mimic the two-pion exchange since the latter is
not well addressed in these models with simple zero-mode quantization.

A straightforward manner to calculate numerically the integrals giving the 
different contributions to the spin-orbit force consists in making the change
in variable ${\bf r}'={\bf x}-{\bf r}/2$ and taking the $NN$ separation 
vector ${\bf r}$ along the $z'$ axis. E.g., with these changes,
the expression of the linear~$\sigma$~model potential $V_{{\rm L}\sigma}$
in Eq.~(\ref{spor}) becomes:
\begin{equation}\label{num}
\displaystyle
V_{{\rm L}\sigma}(r)=-\frac{2\pi}{f_{\pi}^2~r}
\int_0^{\infty} {\rm d}r'~r'\sin^2(F(r'))\xi^2(r')
\int_{-1}^{1} {\rm d}u~u ~\xi^2\left( \sqrt{r^{'2}+r^2+2r r' u}\right) 
\end{equation}
where $u=\hat{\bf r}.\hat{\bf r}'=z'/r'$.
In Fig.2, we plot $V_{{\rm L}\sigma}$, $V_4$, $V_6$ together
with the total sum $V_{{\rm L}\sigma}+V_4+V_6$ as functions of the 
relative distance $r$ between the two nucleons.
As it was the case for the extended Skyrme model~(\ref{La}), $V_4$ and $V_6$
are still repulsive in the case of the model~(\ref{Lanew}). 
On the other hand,
while the nonlinear $\sigma$ model ${\cal L}_{{\rm NL}\sigma}$ 
has a zero contribution to the isoscalar spin-orbit force\footnote{Indeed,
${\cal L}_{{\rm NL}\sigma}$ 
contributes only to the isospin-dependent force~\cite{RN}.}, the linear
$\sigma$ model contributes and with a {\it negative} sign as it can be seen
from the behaviour of $V_{{\rm L}\sigma}$ in Fig.2. This result is in
agreement with the phenomenology~\cite{PB}. The total potential (full
line in Fig.2) is found to be repulsive for intermediate distances 
($r\le 2$fm) and attractive for large distances ($r>2$fm). For short
distances (not shown on Fig.2), neither the product ansatz nor
the effective model should be trusted since this region  
corresponds to processes involving perturbative QCD.
The total potential is attractive at large distances because in this 
region the scalar meson contribution~(\ref{spor}) is the dominant one
as compared to $V_4$ and $V_6$. This can be checked straightforwardly by 
comparing the asymptotic behaviour of each expression in Eq~(\ref{SO}).
We have also displayed in Fig.2 the potential related to the linear 
$\sigma$ model and the sixth-order term only, $V_{{\rm L}\sigma} +V_6$.
We observe from that curve that the attraction starts at about 1.7~fm
instead of the 2~fm in the case of the total potential discussed above.
This result is in some way in agreement with the phenomenological 
isoscalar spin-orbit potential for which
it is well known that scalar-isoscalar and $\omega$ mesons are the mesons
which play the most important role~\cite{PB}. 


Similar results have already been found in Ref.~\cite{KE} where a
different way has been used to introduce a scalar-isoscalar degrees of 
freedom.
In Ref.~\cite{KE}, the scalar-isoscalar meson field is a dilaton, and it 
explicitly couples to the sixth-order term. 
However the contribution to the spin-orbit force coming from the 
sixth-order term was not complete in Ref.~\cite{KE}. Indeed, these authors
ignored the baryon exchange current contribution to the spin-orbit force
in their calculations,
and we have shown in Ref.~\cite{Ab2} that this contribution is significant,
with respect to the direct term~\cite{RS,KE}, 
and should be taken into consideration. It is then legitimate to question 
whether the qualitative results of Ref.~[17] remain valid if the baryon 
exchange current contribution is taken into account.
In this work, by using a different way of including the 
scalar-isoscalar meson in
the Skyrme Lagrangian [14,16], and by considering the {\it entire} 
contribution of the baryon current to the spin-orbit channel, we show that the 
main result of Ref.~\cite{KE} (i.e., an 
attractive $N$-$N$ spin-orbit force at large 
distance from Skyrmions with scalar mesons) is still valid. In this sense,
my results can be viewed as a confirmation of those of 
K\"albermann and Eisenberg~\cite{KE}.


Even though we have obtained an attractive force at large distances
in the isoscalar spin-orbit channel, attraction is still missing 
at intermediate distances, and thus the problem remains unsolved within
 Skyrme-type models. This is not surprising since we
have considered here only a finite-mass scalar field without the other
vector mesons.
However, the result obtained here, namely, the change from a zero
contribution to the isospin-independent spin-orbit interaction to an
attractive one when replacing the frozen $\epsilon$-meson field (nonlinear
$\sigma$ model) with a realistic one (linear $\sigma$ model), is very
encouraging. Indeed, it suggests that the right way to obtain
an attractive isospin-independent $NN$ spin-orbit force within Skyrme-type
models in the framework of the product ansatz is to replace the pion
theory~(\ref{La}) with a realistic effective model including, in addition
to the scalar-isoscalar $\epsilon$ field, all low-lying vector mesons and
taking into account the finiteness of their masses. In a sense, the model
considered here can be seen as a minimal and modest improvement of the pion
Skyrme theory. Similarly, it has been shown in Ref.~\cite{Ka} that the 
model~(\ref{Lanew}) gives rise to 
attraction in the $NN$ central potential but the 
attraction occurs at distances larger than required by phenomenology. The 
problem has been finally solved when the other vectors mesons were 
included~\cite{Ka,KV}. 
For this reason, we believe that in order to cure the problem
of the isoscalar spin-orbit force, one has to consider effective models
which incorporate the first mesonic resonances with finite masses. 
For instance, when considering a finite-mass $\omega$-meson model, 
the $\omega$-field couples directly to the nucleon via the baryon 
current (defined after Eq.~(\ref{La})) and generates the 
common three-pion exchange diagram, contrary to the case of
the local approximation ${\cal L}_6$ ({\it cf.} Eq.~(\ref{La})).
Indeed, in the latter, an unexpected two-pion exchange piece coming from the
baryon exchange current arises in addition to the three-pion piece and
contributes with a positive sign to the isoscalar spin-orbit force
yielding a repulsive interaction~\cite{Ab2}. This problem with the baryon
exchange current is obviously avoided when the sixth order term ${\cal L}_6$
is replaced with a more realistic $\omega$-meson model.

Finally, in addition to considering finite-mass mesons Lagrangians,
we would like to mention a further way that might lead to the desired
attractive isoscalar spin-orbit force within the framework of Skyrme-type
models. It is concerned with the approximation of the product ansatz
configuration~(\ref{pro}). In the latter, the two nucleons are supposed to
keep their spherical shape without deformation even after overlapping.
This is  certainly not true in reality. Thus, while still using the product
ansatz so that we keep benefiting from its simplicity and being able to
perform analytical calculations, we may improve on the 
approximation~(\ref{pro}) 
by allowing the shape of the nucleon to deform when approaching
the other nucleon ({\it cf}.~Ref.~\cite{ROMK} and references therein). 
Calculations of the spin-orbit force within this approach are under 
way~\cite{Ab3}.

\newpage

\begin{figure}[t]
\centerline{\psfig{figure=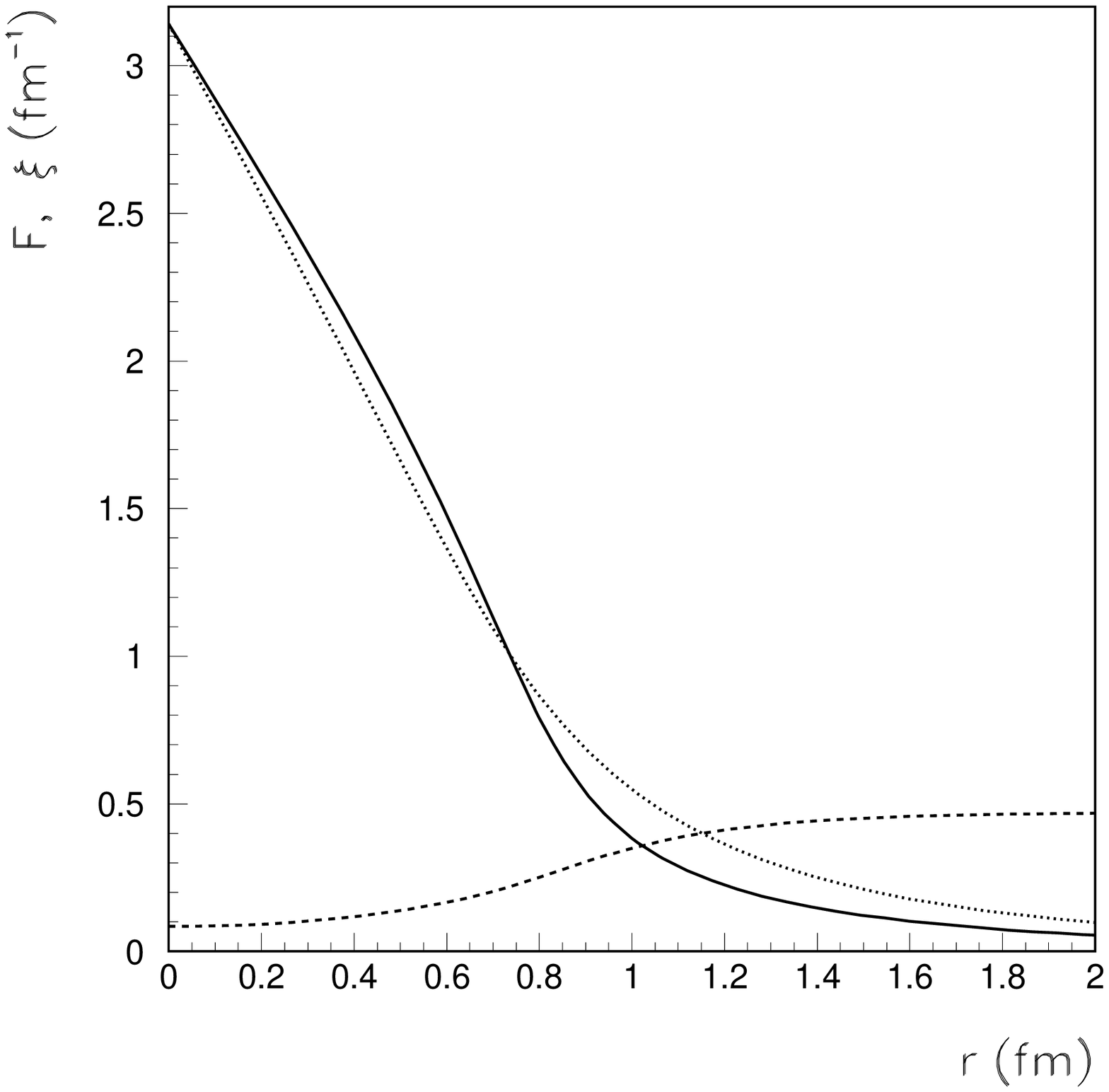,height=8cm,width=14cm}}
\begin{center}{
\parbox{14cm}{ {\footnotesize {\large  FIG.1.}
The chiral function $F$ (full line) and the $\xi$-field (dashed line) 
in~fm$^{-1}$ solutions of Eq.~(\ref{eqs}). The dotted line corresponds
to the chiral field $F$ in the nonlinear model ($\xi=f_{\pi}$).}  }
}\end{center}
\end{figure}

\begin{figure}[h]
\centerline{\psfig{figure=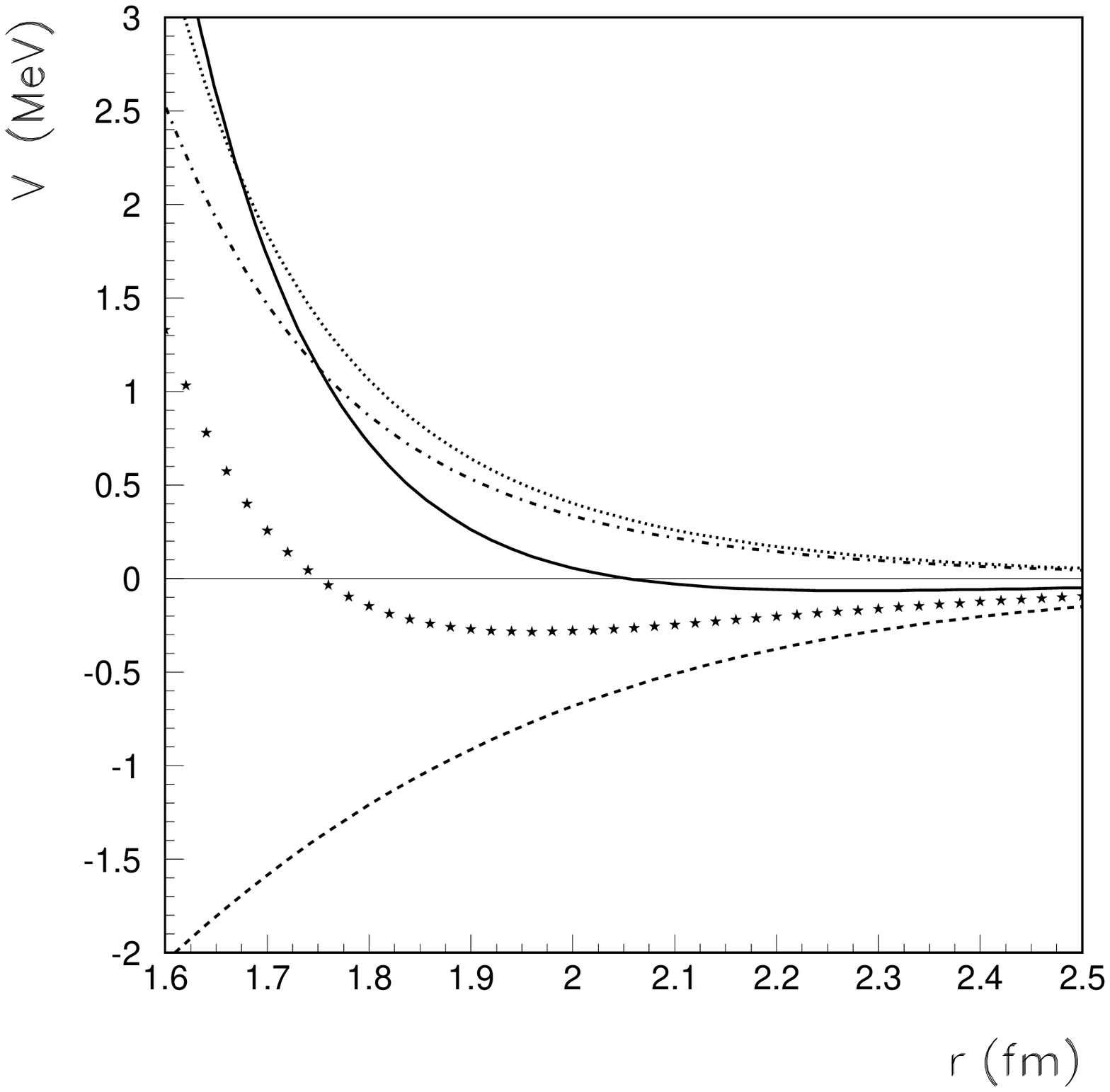,height=8cm,width=14cm}}
\begin{center}{
\parbox{14cm}{ {\footnotesize {\large  FIG.2.}
The potentials $V_{{\rm L}\sigma}$ (dashed line), $V_4$ (dotted line),
$V_6$ (dashed-dotted line), $V_{{\rm L}\sigma} +V_6$ (starry line) and
$V_{{\rm L}\sigma} +V_4+V_6$ (full line) 
with respect to the relative $NN$ distance $r$.} }
}\end{center}
\end{figure}


\begin{thebibliography}{99}
\bibitem{Sk} T. H. R. Skyrme, Proc. Roy. Soc. {\bf A260}, 127 (1961).
\bibitem{Wi} E. Witten, Nucl. Phys. B {\bf 223},  422 and 433 (1983). 
\bibitem{ZB} I. Zahed and G. E. Brown, Phys. Rep. {\bf 142}, 1 (1986).
\bibitem{ANW} G. S. Adkins, C. R. Nappi, and E. Witten, Nucl. Phys. B
{\bf 228}, 552 (1983).
\bibitem{Mou} B. Moussallam, Ann. Phys. (N.Y.) {\bf 225}, 264 (1993).
\bibitem{JJP} A. Jackson, A. D. Jackson, and V. Pasquier,
Nucl. Phys. A {\bf 432}, 567 (1985).
\bibitem{Jac} A. Jackson {\it et al}, Phys. Lett. B {\bf 154}, 101 (1985).
\bibitem{AM} A. Abada and H. Merabet, Phys. Rev. D {\bf 48}, 2337 (1993).
\bibitem{autho} E. M. Nyman and D. O. Riska, Phys. Lett. B {\bf 175}, 392 
(1986);\\ 
D. O. Riska and K. Dannbom, Phys. Scr. {\bf 37}, 7 (1988);\\ 
T. Otofuji {\it et al}, Phys. Lett. B {\bf 205}, 145 (1988).
\bibitem{WA} R. D. Amado {\it et al}, Phys. Lett. B {\bf 314}, 159 (1993); 
Phys. Lett. B {\bf 324}, 467 (1994);\\
B. Shao {\it et al}, Phys. Rev. C {\bf 48}, 2498 (1993);   
Phys. Rev. C {\bf 49}, 3360 (1994);\\
A. Abada,  ``{\it The isospin independent spin-orbit force in the extended 
Skyrme model}'', hep-ph/9401341, unpublished.
\bibitem{PB} M. Lacombe {\it et al}, Phys. Rev. C{\bf 21}, 861 (1980);\\
R. Machleidt, K. Holinde and Ch. Elster, Phys. Rep. {\bf 149}, 1 (1987).
\bibitem{RS} D. O. Riska and B. Schwesinger, Phys. Lett. B {\bf 229}, 339 
(1989). 
\bibitem{Ab2} A. Abada, J. Phys. G {\bf 22}, L57 (1996).
\bibitem{Ka} D. Kalafatis, hep-ph/9406410, thesis, Univ. Paris XI.
\bibitem{Ike} K. Iketani, Kyushu University preprint 84-HE-2 (1984),
unpublished.
\bibitem{KV} D. Kalafatis and R. Vinh Mau, Phys. Rev. D {\bf 46}, 3903 (1992)
and references therein.
\bibitem{KE} G. K\"albermann and J. M. Eisenberg, Phys. Lett. B {\bf 349},
416 (1995).
\bibitem{Mk} M. C. Birse, J. Phys. G {\bf 20}, 1287 (1994).
\bibitem{Sk2} T. H. R. Skyrme, Nucl. Phys. {\bf 31}, 556 (1962).
\bibitem{NW} N. R. Walet, Nucl. Phys. A {\bf 586}, 649 (1995) and references 
therein.
\bibitem{Nym} E. M. Nyman, Phys. Lett. B {\bf 162}, 244 (1985).
\bibitem{RN} D. O. Riska and E. M. Nyman, Phys. Lett. B {\bf 183}, 7 (1987).
\bibitem{ROMK} A. Rahimov {\it et al}, Phys. Lett. B {378}, 12 (1996).
\bibitem{Ab3} A. Abada, work in progress.
\end{thebibliography}
\end{document}